\documentclass[12pt]{article}
\usepackage[utf8]{inputenc}
\usepackage{natbib}
\usepackage{float}
\usepackage{hyperref}
\title{Response of  Indian East coast upwelling  to river runoff in an OGCM}
\author{C. P. Neema and P. N. Vinayachandran \\ Centre for Atmospheric and Oceanic Sciences and  Divecha Centre for Climate Change\\ Indian Institute of Science, Bengaluru, India}
\date{\today}
\usepackage{graphicx}
\usepackage[paperheight=29.7cm,paperwidth=21cm,textwidth=17cm]{geometry}
\usepackage{setspace}
\usepackage{comment}
\begin{document}

\maketitle{}
\begin{abstract}


Along the east coast of India (ECI), local winds are oriented alongshore during the summer monsoon, favouring coastal upwelling. The East India Coastal Current (EICC)  during this period converges at  17$^{\circ}$N  with northward flow from the south and southward flow from the north.   A high-resolution  Ocean general circulation model (OGCM) which includes runoff from major  rivers  (RIVER) simulates the upwelling along the east coast of India during the summer monsoon reasonably well.  The model simulation  shows that  the upwelling  increases towards south, where salinity influence is lesser compared to northern parts. The presence of rivers suppresses upwelling,  the depth of the source of subsurface water becomes shallow  from 100 to 60 m, and reduces the SST cooling by 0.5$^{\circ}$C owing to the increase in stratification of the water column.

\end{abstract}

\section{Introduction}

Coastal upwelling is a natural oceanographic process that involves the movement of subsurface cold nutrient-rich water  to the surface altering the  biological productivity  of the region.  It occurs as a result of the ocean's surface water being pushed away from the coast by the wind, to be replaced by water from beneath. 
 Coastal upwelling  occurs when winds blow parallel to the coast.  Supply of nutrients and minerals from the subsurface ocean  to  the upper layers  fuels the growth of phytoplankton,  supports the food chain, and promotes fisheries.  Most of the major upwelling systems are located along  the eastern boundaries of the ocean; examples are California, Peru-Chile, Portugal-Northwest Africa, and Southwest Africa, where the prevailing winds are equatorward \citep{kampf2016upwelling}. Seasonal upwelling occurs along Somalia, Sumatra, Yemen, South China Sea, southwest coast of India, southern tip,  east coast of India (ECI),  and the Sri Lankan dome region \citep{vinay2021, hood2017} despite the eastern or western orientation of the coast.

Along the ECI, local winds  are oriented alongshore and northwards favouring upwelling during the summer monsoon \citep{shetye1991wind, RAO1995, thushara2016}. The winds become favourable for upwelling during the pre-monsoon itself \citep{rao1989} and upwelling has been reported along Vishakapatanam coast from February which  continues till July \citep{rao1986upwelling}.  Upwelling was found to extend spatially offshore till 50 km from 13$^{\circ}$ to 18$^{\circ}$N along the east coast \citep{rao2002spatial}.  
The barrier layer formed in the BoB caused by the increased stratification due to river influx can inhibit the upwelling in the region \citep{behara2016}.    \citet{thushara2016} showed that the upwelling was impacted by the river water  during the later part of the monsoon season along the ECI. In addition to the freshwater discharge and alongshore wind stress, mesoscale eddies, remote winds from the equator, and wave propagation from the interior of the bay modulate  coastal upwelling in this region \citep{vinay2021}.   \citet{AMOL2020} and \citet{thushara2016} have examined the impact of upwelling on the chlorophyll blooms in the northwestern bay and found that an increase in stratification limits the vertical supply of nutrients in the upwelling regions along the east coast. \citet{thushara2016} cited upwelling and a minor role from advection as the cause for the transport of nutrients from the coast to the offshore in the northern bay.  Recently \citet{DEY2023} noticed active and break phases in upwelling using HF radar data. They observed that, despite the Ekman transport showing upwelling favourable conditions from April to August, SST and currents exhibited upwelling until mid-June and then ceased to exist as a result of two processes; stratification by freshwater and the arrival of a downwelling Kelvin wave. \cite{ray2022} similarly reported that the upwelling Kelvin wave during January to April triggered coastal upwelling along the northeast coast of India and it was suppressed by a downwelling Kelvin wave during April-August. The propagation of the upwelling and  downwelling Kelvin waves weakens along the western boundary of the bay  due to the prevailing circulation in this region.  The poleward flowing EICC and a clockwise eddy in the head bay seems to play a significant role in this weakening \citep{RAO2010}.   

With the advance of the summer monsoon, freshwater discharge into the BoB increases, lowering the salinity in the region. A decrease or increase in rainfall or river discharge can cause significant changes in the surface salinity of the bay.  Stratification induced by low salinity water can have important implications on the ECI upwelling as the rivers flow equatorward in this region.    A numerical modeling study by \citet{john92}, forced  with southward transport of water of Hugli and Mahanadi rivers showed the  suppression of  upwelling north of  Vishakapatanam by the low salinity plume and this effect  decreased south of it. 

\citet{thushara2016} studied the chlorophyll blooms in the western bay during upwelling through  model experiments. In particular, they examined the bloom dynamics associated with stratification induced by river flux and wind transports.  As the upwelling process diverts the plume away from the coast,  low salinity by river flux had little or no impact on the blooms. There have not been any attempts to   quantify the role of rivers in upwelling in this region.  In this study,  we have used a high-resolution (nearly 13 km) ocean model to simulate the features of upwelling, 
and  the role played by river runoff  in modulating the upwelling along the ECI during the summer monsoon using  ocean model experiments.  We find that the river-induced stratification  plays an important role in suppressing upwelling in the northern section of the east coast.   The paper is organised as follows.  Section 2 provides a brief overview of the model simulations, data, methodology, and validation. In sections 3 and 4, we discuss how the model simulates upwelling in the BoB and the effect of rivers on it. A description of river-induced stratification is provided in Section 5, and a conclusion and summary are given in Section 6.

\section{Model, data, and methodology}

A global  ocean general circulation model (MOM version 5.1) with a horizontal resolution 1/8$^{\circ}$ (approximately 13 km in the Bay of Bengal) has been used for the present study \citep{vinayachandran2022fate}.  The model has 42 vertical levels with a resolution of 5 m in the upper 55 m and  the resolution decreases exponentially with depth. The horizontal grid is tripolar and the topography is based on \cite{sindhu2007}.  The vertical mixing scheme is KPP \citep{large1994} with the bulk Richardson number set to 0.3. For horizontal mixing, a combination of Laplacian and biharmonic friction with a Smagorinsky mixing coefficient of 0.01 and 0.1 respectively, has been used. CORE II \citep{large2009global} data sets are used for long wave, short wave, air temperature, specific humidity, winds, snow, and rainfall data.  Chlorophyll data, used for the parametrization of shortwave penetration in the ocean, is monthly and was acquired from SeaWiFS.   

Rivers entering the ocean are distributed horizontally and vertically by the prevailing current systems in accordance with the thermohaline circulation.   In the Indian Ocean, river runoff and its advection lead to large inter-annual variability in salinity near the river mouths and in the eastern equatorial Indian Ocean \citep{vinayravi2009}.  
  \citet{kurian2006}  used the spreading method to incorporate river discharge into the modular ocean model; runoff from each river is spread over a horizontal area proportional to its discharge rate over time. In addition to offering the flexibility of choosing the river spreading area, this method is found to be very successful in simulating the surface salinity of the Indian Ocean  \citep{kurian2006, kurian2007, vinayachandran2012, behara2016}. Hence, we have employed a similar method to include the global river discharge in the high-resolution model \citep{vinayachandran2022fate}.  
 The exact location of the river mouth is identified and fixed in the model, for each river. Accordingly, the extent of each river in the ocean is decided based on the magnitude of the river discharge during each month. Spreading zones are chosen such that water supplied to each grid box does not exceed 0.05m/day, a value recommended \citep{Griffies2004}, to maintain model stability. Thus, the area over which the discharge of a river is distributed depends on the discharge volume. Vertically, the river is discharged in the upper 10 m, which is predetermined  in the model such that each layer receives a fraction of the total discharge.   The inter-annual monthly river runoff \citep{dai2017} available from 1900 to 2014, is used for river discharge in the model. However, large data gaps  existed for many rivers for several years during this period.  Hence a monthly climatology was prepared from the available years for each river system. Based on the minimum thickness, global scale, and vertical resolution, the upper 10 m has been chosen for the distribution of  runoff.  

The model spin-up was carried out using initial conditions  for 175 years starting from a state of rest.  Additional experiments branch off for a further 40 years  from this spinup.  
The climatological run (RIVER) with all the major rivers included and a designed experiment where  all the rivers (NORIVER) were switched off are utilised to study the impact of rivers on the global ocean. In the present study, we utilise the 40$^{th}$ year, monthly simulations confined to the BoB.  The difference plot refers to RIVER minus NORIVER simulation.

\begin{figure}[ht]
\centering
\includegraphics[width=1.\columnwidth]{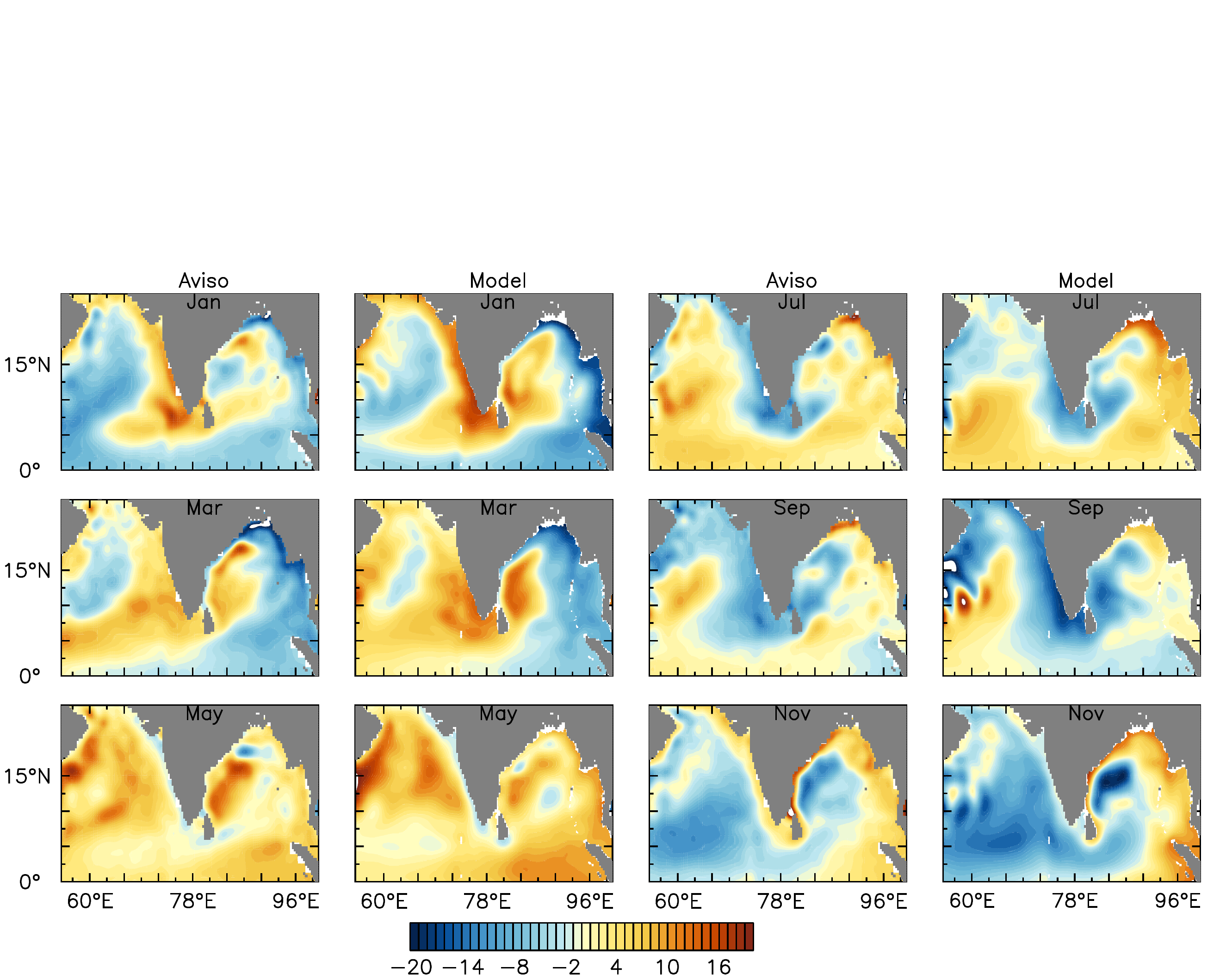} 
\caption{Bimonthly climatology maps of SSH (cm) anomalies from the model (RIVER run) and AVISO.}
\label{SSHriver}
\end{figure}

\begin{figure}[ht]
\centering
\includegraphics[width=1.\columnwidth]{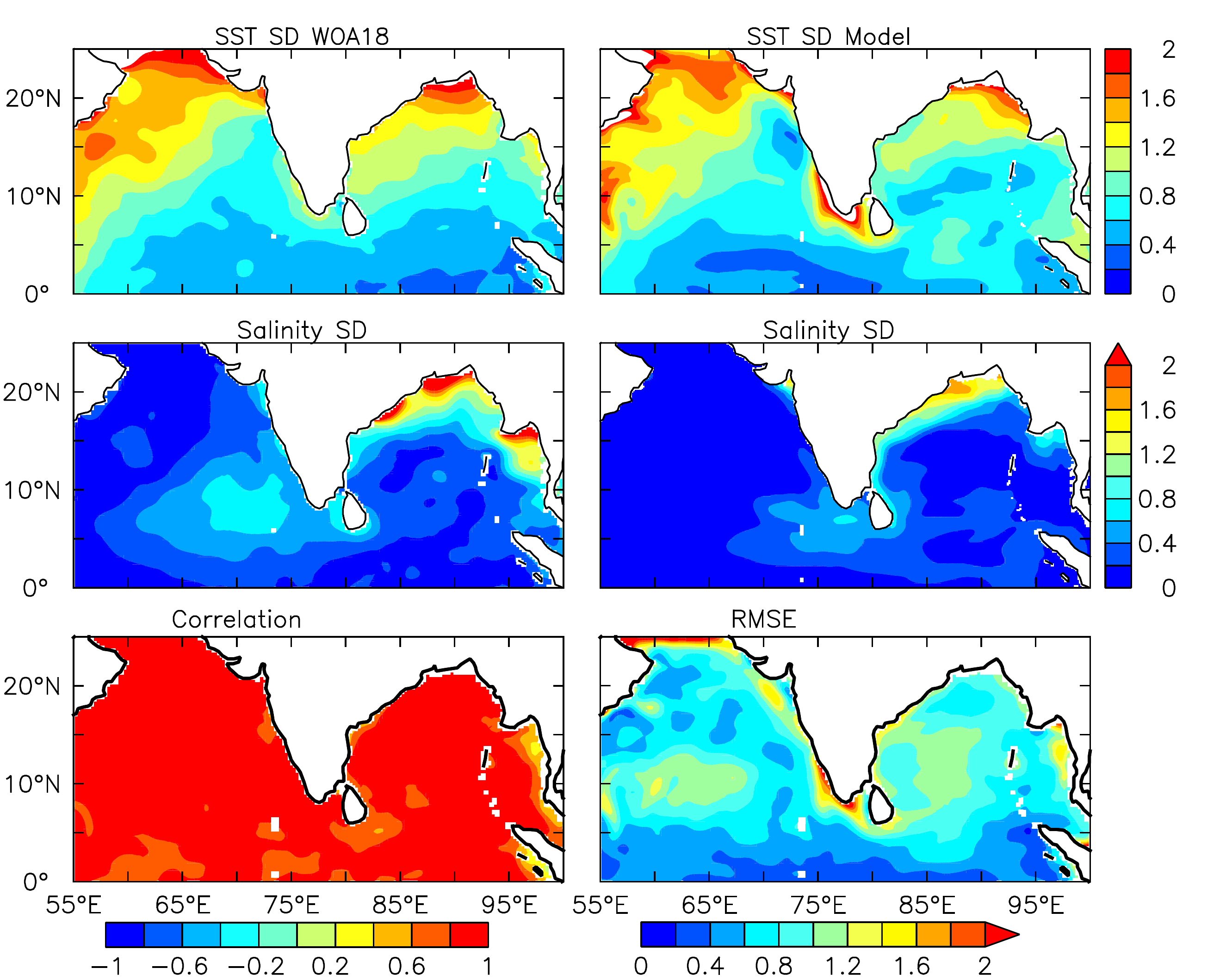} 
\caption{Annual standard deviation for SST (upper panel ), and Salinity (middle panel) from WOA18 (left panel) and model (right panel).  Bottom panel shows the correlation (left) and RMSE (right) between the model and WOA18 SST. The model results are from the RIVER run.}
\label{sd_mod_obs}
\end{figure}

   Model Sea surface height (SSH) anomaly and AVISO (\url{https://www.aviso.altimetry.fr/en/data.html}) SSH anomaly shows  good agreement. Major features of the  SSH anomaly (Figure \ref{SSHriver}) are simulated well by the model. The Lakshadweep high and low during winter and summer respectively are reproduced by the model and it reproduces the cyclonic low  and anticyclonic high in the BoB also remarkably well. 
Validation of the model for SST, salinity, and currents can be found in \citet{vinayachandran2022fate}.
The SST variability in the BoB, low salinity plume, and its circulation along the ECI are reproduced by the model reasonably well (see Figure 1 and Figure 2 from \citet{vinayachandran2022fate}). The major variability in  SST occurs in the northern BoB, southwest coast of India, and northern Arabian Sea and shows a good comparison with observations (Figure \ref{sd_mod_obs}).   Though the salinity standard deviation  from the model captures  the prominent patterns of variability such as those near the major discharge region in the BOB and the southwest coast of India, the magnitude seems weaker compared to observations. 
The correlation of model SST with WOA18 \citep{boyer2018world} SST is above 0.8 over most parts of the basin (last panel, Figure \ref{sd_mod_obs}). The root mean square error (RMSE)  estimates are higher along the southwest coast of India compared to observation.  The rest of the  regions show errors less than 1$^{\circ}$C. In general, the model simulates the general circulation features and thermohaline variability of the north Indian Ocean (Figure 1 and Figure 2. from \citet{vinayachandran2022fate}) similar to observations and hence can be used as a tool to study the ocean dynamics in the region.  



\section{BoB Upwelling Simulation in the Model}

\begin{figure}[ht]
\centering
\includegraphics[width=0.8\textwidth]{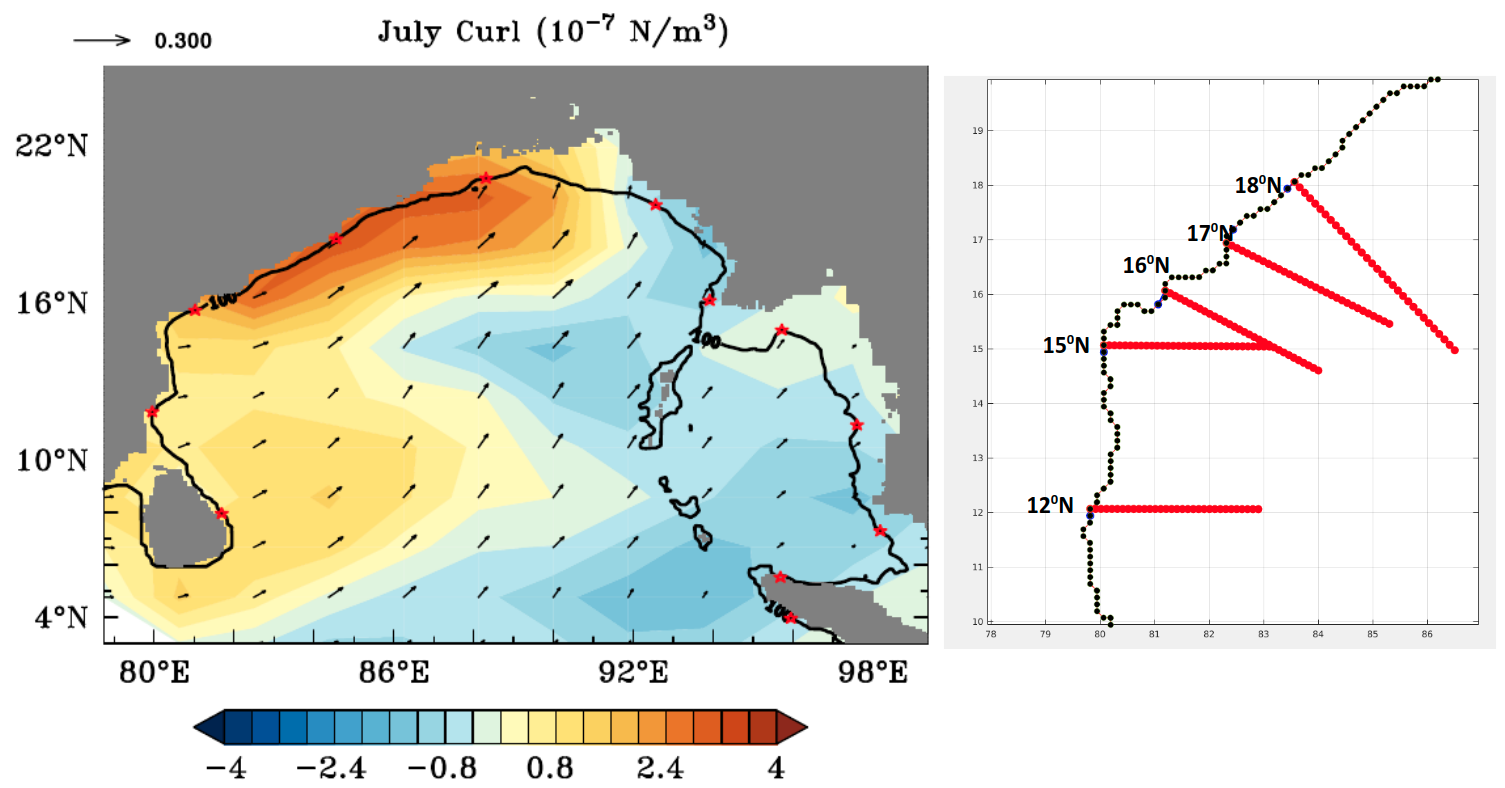} 
\caption{Left panel. Climatological wind stress (vectors, m/s) and curl (shading)  calculated from CORE II \citep{large2009global} 6 hourly data for the month of July. Contour represents the  100 m isobath used in the model.   Right panel. The coastal stretch (black line)  considered for Figure 6, Figure 12 and sections    (red lines) for Figures 7, 9, and 10.  The latitude of the section at the coast is marked in the figure.}
\label{wind_stress}
\end{figure}

\begin{figure}[ht]
\centering
\includegraphics[width=1.\columnwidth]{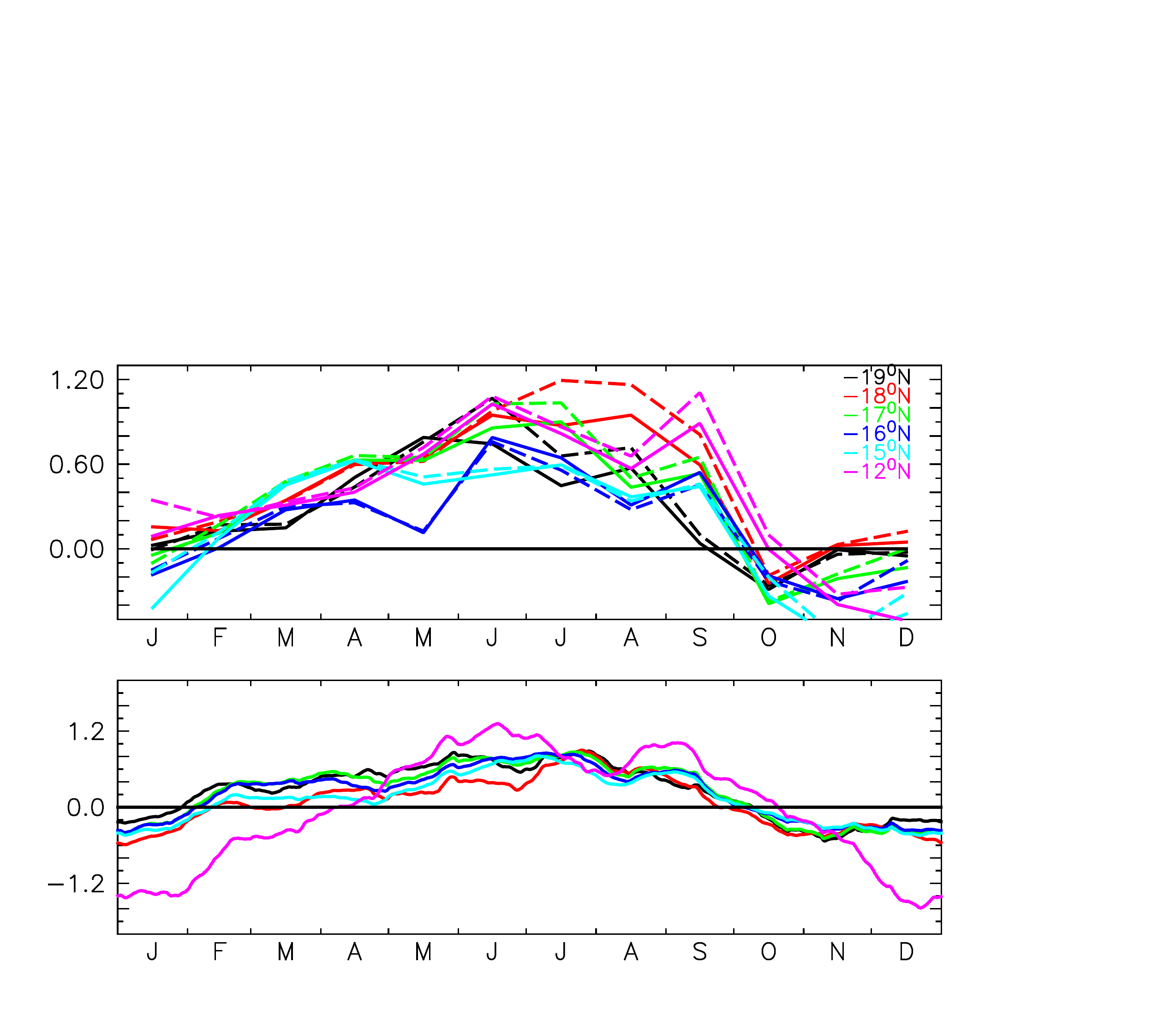} 
 \vspace{1pt} 
\caption{Upwelling Index (UI$_{sst}$) computed from SST difference from the coast and 1-degree latitude away from the coastal point along the east coast of India (upper panel). Solid curves represent the UI$_{sst}$ from RIVER and dashed ones  from  NORIVER respectively. Upwelling index computed from alongshore wind stress (UI, bottom panel).}
\label{UI_sst}
\end{figure}

The southwest monsoon winds are  nearly   parallel to the ECI with high strength during June - August period (only July shown, Figure \ref{wind_stress}) and thus favourable for Ekman type of upwelling. The curl of the  wind stress  is also positive and favourable for upwelling  during the May-September period (only July shown, Figure \ref{wind_stress}).  The cyclonic  curl strengthens during June-July along the east coast from around 14$^{\circ}$N to 20$^{\circ}$N  and  weakens by August.  To quantify Ekman upwelling,  the upwelling index (UI) is calculated  using the formula  UI=$\tau{_a}/\rho f$ \citep{bakun1975}, where $\tau{_a}$ is the alongshore wind stress, $\rho$ is the density of water, and {$f$} is the Coriolis   force. Positive values  of UI indicate upwelling (Figure \ref{UI_sst}).

 As a proxy to represent the upwelling signatures from the ocean, we have calculated the SST index (UI$_{sst}$) from model simulations. The difference of SST between the coast  and at a location  100 km  away from the coast at the same latitude following \citet{patel2020} is considered here as the SST index for upwelling. Thus a positive value of UI$_{sst}$ indicates cooling near the coast  which can be used to  quantify the upwelling that is realistically taking place in the ocean. 
 
 Temporal variability of UI calculated for the ECI at the selected location from 19$^{\circ}$ to 12 $^{\circ}$N are presented in Figure \ref{UI_sst} (second  row , monthly smoothed).  The wind-driven upwelling starts  in February and is positive  (maximum of 1.2 $m^{2}s^{-1}$) for all locations along the coast till September except at 12$^{\circ}$N, the southernmost location (purple line, upper and lower panel, Figure \ref{UI_sst}),  which becomes positive from May onwards and is higher than that at the other location for the summer monsoon period. 
     The UI$_{sst}$ (upper panel, Figure \ref{UI_sst}) is  positive (maximum  of 1.2 $^{\circ}$C) suggesting  upwelling  from February till September. High positive values that begin to appear in June  indicate the presence of upwelling during the summer monsoon and it decreases rapidly during October.

\begin{figure}[ht]
\centering
\includegraphics[width=.7\textwidth]{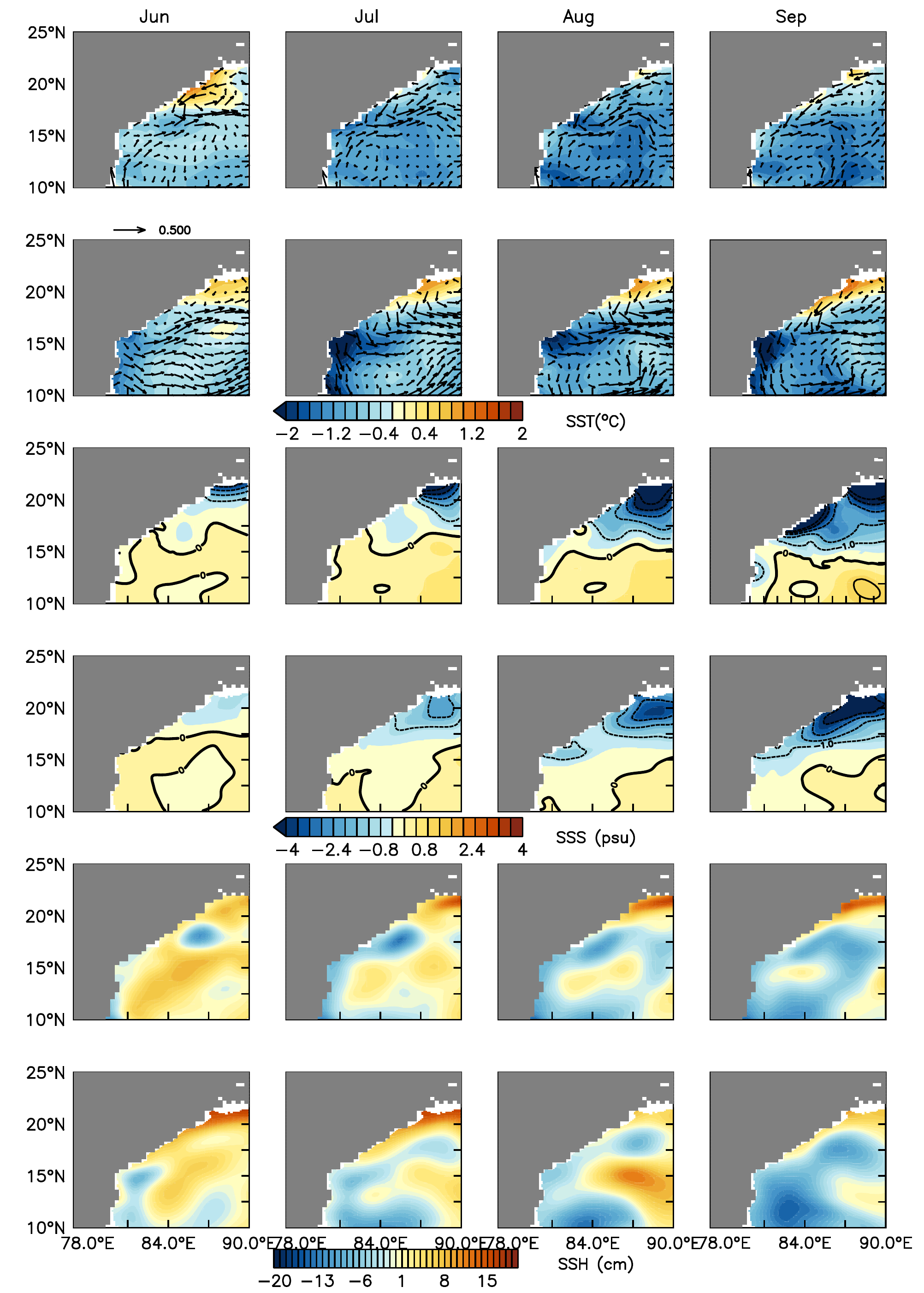} 
\caption{Row 1. Monthly mean currents (vectors, m s $^{-1}$) from OSCAR and  the difference in  SST (shading) from that for the month of May from WOA18 (top panel). Row 2.  Same as in Row 1 except that current and SST differences are from the model. Row 3.  The difference in  WOA18 salinity from that for the month of May. Row 4. Same as in Row 3 except that differences are from Model. Row 5. SSH from AVISO and Row 6. SSH from Model for the summer monsoon period.  The model results are from the RIVER run.}
\label{sst_ssh_winds}
\end{figure}

 The EICC as seen in the OSCAR data set (first row, Figure \ref{sst_ssh_winds}) flows northward upto about 17$^{\circ}$N during the month of June and  then turns offshore. The inshore region of this offshoot shows a cyclonic eddy with equatorward flow close to the coast. With the progress in the summer monsoon, the northward flow weakens and the equatorward flow gets established along the northern half of the coast. The cyclonic eddy persists during June - August but weakens in  September. The model EICC (second row, Figure \ref{sst_ssh_winds}) is consistent with the OSCAR.  The equatorward flow in the northern section of the coast and the poleward flow seaward of it are well reproduced by the model. The major difference is seen in the model EICC in the southern sector of the coast which flows  more in the offshore direction than in the OSCAR data set.  The model simulation offers a good comparison with the SSH anomaly from AVISO (5$^{th}$ and 6$^{th}$ row, Figure \ref{sst_ssh_winds}).  SSHA is higher in the entire western bay during June which changes signs towards the end of the season.  A  low sea level corresponding to the cyclonic eddy is noticed  around 18$^{\circ}$N near the coast. The formation of this eddy occurs a while later in the model but the alternating bands of low and high sea levels corresponding to the circulation (5$^{th}$ and 6$^{th}$ row, Figure \ref{sst_ssh_winds}) is in good agreement with the altimetry. 

\begin{figure}[ht]
\centering
\includegraphics[width=0.8\columnwidth]{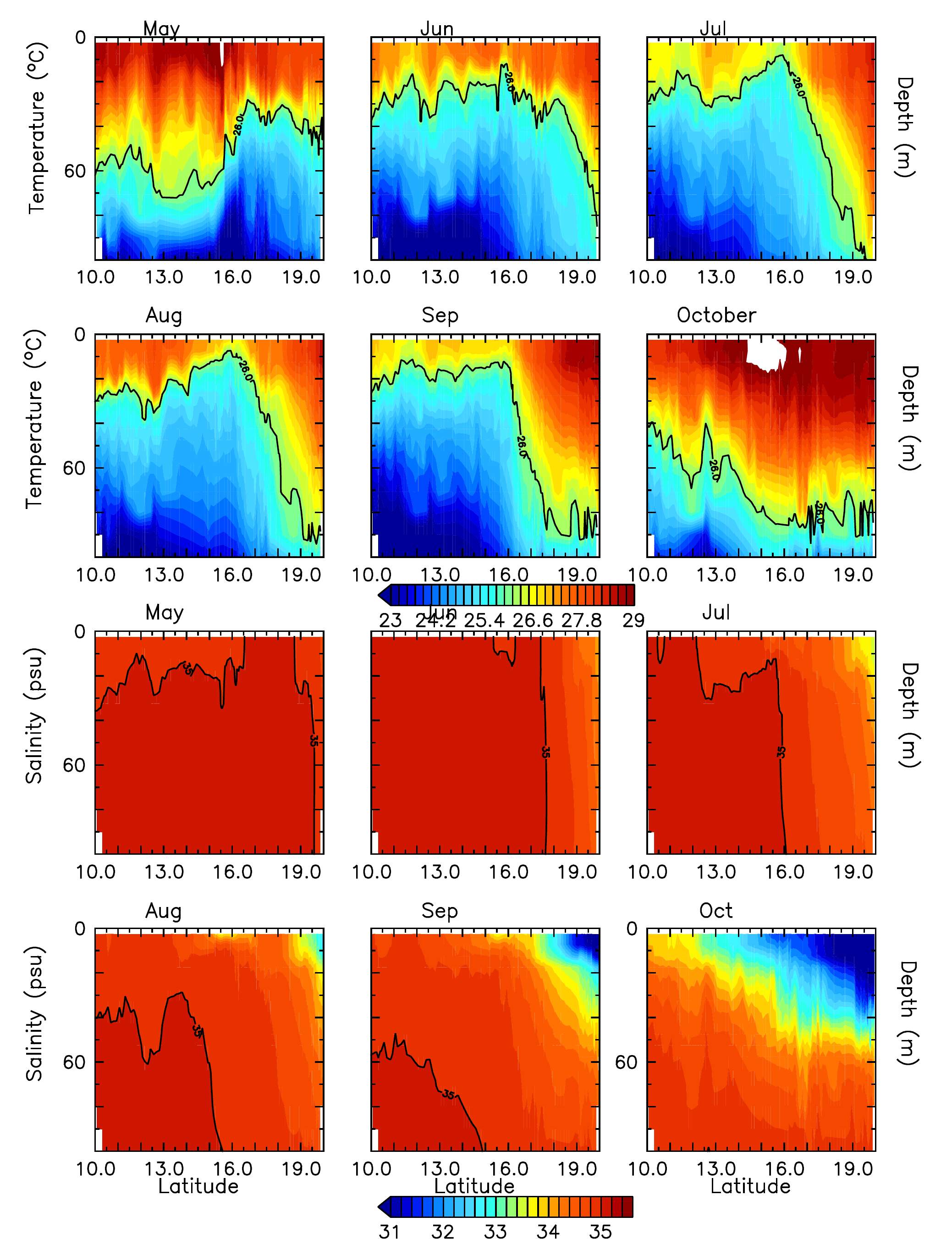} 
\caption{Vertical structure of Temperature (first and second rows) and Salinity  (third and fourth rows) along the 100 m isobath of the part of the coastline shown in the right panel of Figure 3 from May-October. The contour marked on the temperature plot is 26$^{\circ}$C and on salinity is 35 psu. All results are from the RIVER run.}
\label{may_oct_lat_depth}
\end{figure}

 In order to show the cooling associated with upwelling along the ECI during the summer monsoon months of June, July, August, and September, the SST from the model (RIVER) simulation and WOA18 are utilised and their difference from May are shown (first two rows, Figure \ref{sst_ssh_winds}).  
In June, close to the coast, northwestern bay (till 18$^{\circ}$-20$^{\circ}$N) shows warming (0.5-1$^{\circ}$C)  while the southwestern bay cooling with respect to May (Figure \ref{sst_ssh_winds}).  Both model and data show warming in the northern part of the coast (north of 18$^{\circ}$N) and cooling in the southern part  due to the monsoon. In July and August, the northern bay continues to show warming compared to May  which becomes more pronounced and extends southward in September.  The SST cooling  in the south increases during July and August (1-2$^{\circ}$C).  Thus both the model and observation imply an  increase in SST cooling south of 18$^{\circ}$N  during the summer monsoon associated with upwelling.  The strength of upwelling shows considerable latitudinal variation along the coast within the season with a noticeable impact on the vertical structure. Figure \ref{may_oct_lat_depth} shows the latitude depth section of temperature and salinity  along the ECI (along the 100m isobath)  from 10$^{\circ}$- 20$^{\circ}$ N from  May to October.  Prior to the summer monsoon, the coastal region to the south of 16 $^{\circ}$N exhibits higher temperatures  (29$^{\circ}$C) at the surface. With the start of the summer monsoon in June, the isotherms rise by about 30 to 50 m compared to their pre-monsoon position in these southern parts.   In the north, however, there is a deepening of similar magnitude during the same period. SST is also maintained above 28$^{\circ}$C in the north and drops  to 27$^{\circ}$C in the south during the summer monsoon. Another interesting feature noticed here is the sudden increase in the upper layer temperatures (28$^{\circ}$C) from July to August south of 15$^{\circ}$N. This corresponds well with the UI and  UI$_{SST}$, which also shows a slight decrease in its magnitude during August (Figure \ref{UI_sst}).   Local alongshore winds apparently slow down during August (Figure not shown).  Breaks in upwelling were noticed in a recent study using high-frequency HF radar data \citep{DEY2023} where they associated it with the local wind variability. 

 Beginning in June, a  low salinity plume ($<$ 2 psu ) expands southward from the northwest bay and it reaches 16$^{\circ}$N in September (3$^{rd}$ and 4$^{th}$ row, Figure \ref{sst_ssh_winds}).
 The pattern is similar in the model with the low salinity values  slightly less  than that of observations. 
 In contrast, there is an increase in salinity to the south of 18$^{\circ}$N which can be attributed to the upwelling of high salinity water near the coast \citep{shetye1991wind}. 
 A freshwater plume progressively moves southward along the coast along with an increase in the vertical extent (3$^{rd}$ and 4$^{th}$ row, Figure \ref{may_oct_lat_depth}).  
  The salinity increases near the surface due to upwelling in June but decreases thereafter due to the southward flow of freshwater.  Higher temperatures are maintained at the northern parts of ECI, where low salinity values are found, along with  thinning of the mixed layer (Figure not shown).   By the end of October, the low saline values (less than 34 psu)  spread along the entire coast.

\begin{figure}[ht]
\centering
\includegraphics[width=1\columnwidth]{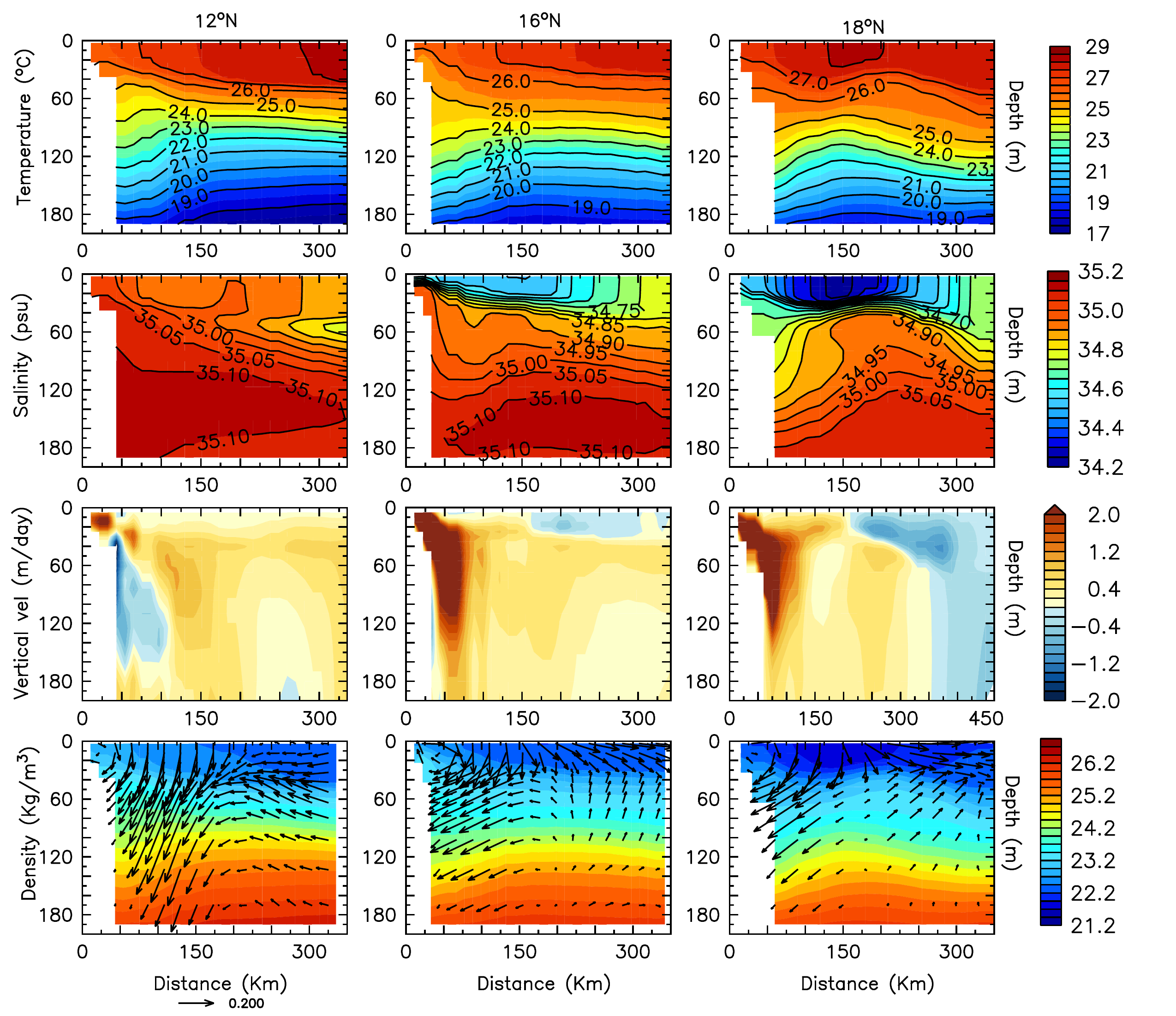} 
\caption{Vertical structure of Temperature (1$^{st}$  row), Salinity  (2$^{nd}$  row),  and Vertical velocity (3$^{rd}$ row) for the month of July for the sections marked at 12$^{\circ}$N, 16$^{\circ}$N and 18$^{\circ}$N in the right panel of Figure 3. All results are from the RIVER run.}
\label{TSwt}
\end{figure}

 
  Cross-shore vertical section (Figure \ref{TSwt}) taken perpendicular to the  coast till 300 km in the  horizontal  and 200 m   vertically was selected to illustrate the variations  in temperature, salinity, vertical velocity, density, and currents during upwelling  at three different locations  along the ECI; one each in the south (12$^{\circ}$N),  center (16$^{\circ}$N), and north (18$^{\circ}$N). In the upper 60 m, isotherms are inclined upward near the coast, indicating upwelling  (1$^{st}$  row, Figure \ref{TSwt}). Temperatures cool maximum to the south (26$^{\circ}$C, at 12$^{\circ}$ and 16$^{\circ}$N ) compared to the north  (27$^{\circ}$C, 18$^{\circ}$N) in the upper layers close to the coast.  Downwelling is noticed below 60 m near the coast with the isotherms doming downwards consistent  with observations \citep{shetye1991wind} in this region.  The northernmost station shows the lowest salinity (18$^{\circ}$N, 34 psu) with values increasing southward.  Even though low salinity water is present in the upper layers, upwelling tends to bring saltier water from beneath to the surface  near the coast  (second row, Figure \ref{TSwt}).
 Vertical velocity is upward (positive) (3$^{rd}$ row, Figure \ref{TSwt}) at all three locations in the upper layers especially near the coast.  
 Density decreases offshore  (22 Kg/m$^{3}$, 4$^{th}$ row, Figure \ref{TSwt}) whereas upwelling brings comparatively denser waters to the top near the coast at all the locations. The cross-shore deflection of currents to the right is apparent up to a distance of 50-100 km (vectors, 4$^{th}$ row, Figure \ref{TSwt}) at the surface.  These currents turn  clockwise  with depth and reverse at a  depth of about 30-40 m completing a clockwise circulation.    


  In summary,  the high-resolution ocean model realistically simulates the upwelling that occurs along India's east coast through the summer. The alongshore wind stress-based UI showed upwelling starting from February and lasting till September.  The model-based UI$_{sst}$ displayed the model's ability to capture the upwelling response from the ocean.  
  The surface winds favouring upwelling along the east coast of India produce a differential upwelling pattern north and south of 18$^{\circ}$N in the western bay. This is in accordance with the currents  which have a poleward component south of 18$^{\circ}$N  and equatorward  north of it.    
  The SST cooling associated with upwelling increases as one moves south, where the effects of the low salinity plume from the northern bay are less prominent and high salinity water (35 psu) slopes up. The western bay cools more to the south of 18$^{\circ}$N and warms north of it with respect to the month of May.  The warming becomes more pronounced and extends southward by September.  There is a break in the strength of upwelling during August, south of 16$^{\circ}$N, when the temperature starts increasing corresponding to the weakening of the wind stress curl in the region.  The upwelling is noticed till a  depth of 60 m in the vertical and through this process, higher salinity denser water reaches the surface.  The cross-shore response of  current \citep{DEY2023}, typical of upwelling is evident in the region upto a distance of about 50-100 km offshore.


\section{Impact of river runoff on upwelling}

\begin{figure}[ht]
\centering
\includegraphics[width=.8\textwidth]{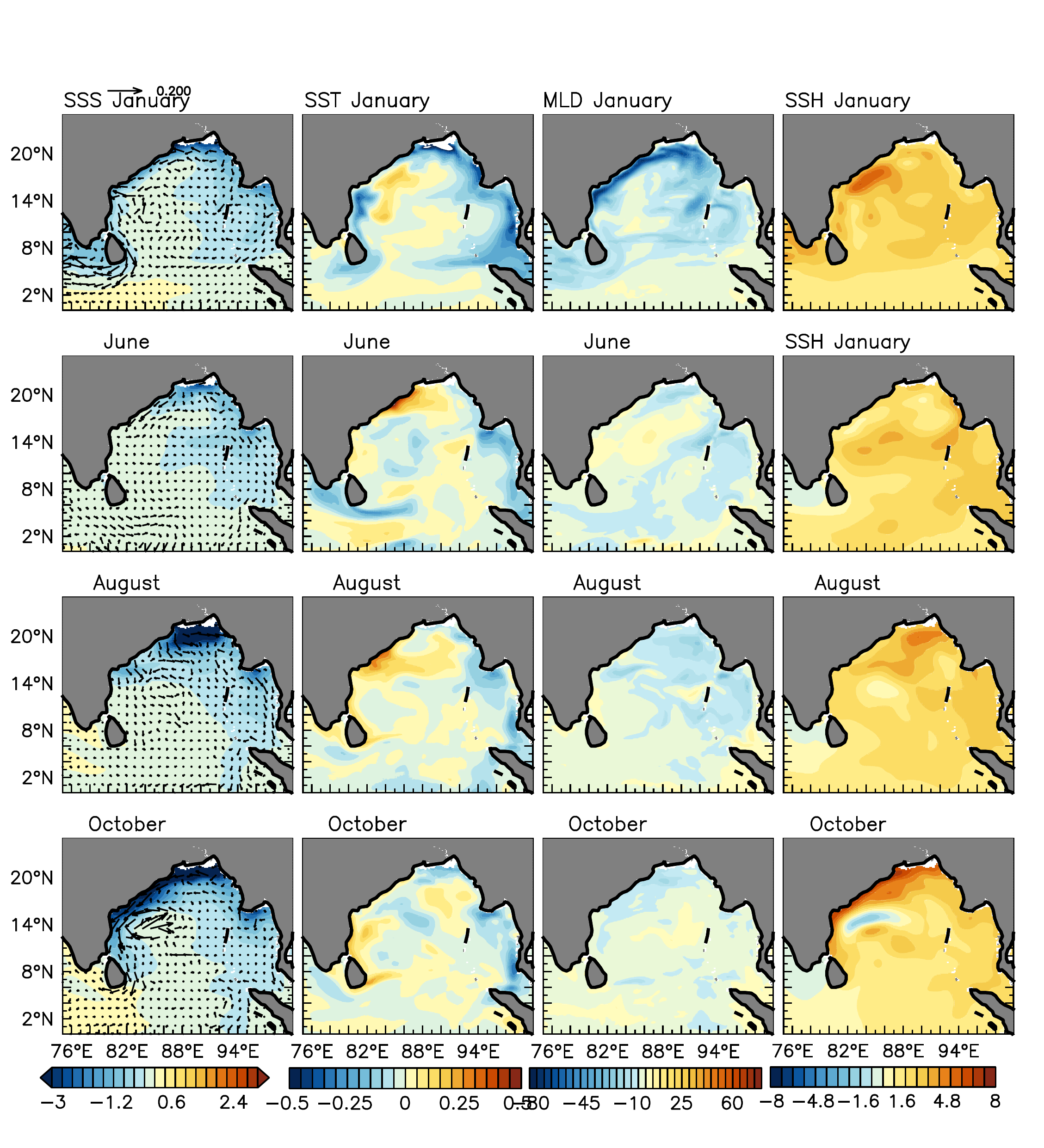} 
\caption{Difference between RIVER and NORIVER for SSS (psu, shaded, 1$^{st}$ column)  and currents (m/s, vectors, 1$^{st}$ column), SST ($^{\circ}$C, 2$^{nd}$ column), MLD (m, 3$^{rd}$ column) and SSH (cm, 4$^{th}$ column) for January (1$^{st}$ row),  June (2$^{nd}$ row),   August (3$^{rd}$ row),  and October (4$^{th}$ row).}
\label{allvar_diff}
\end{figure}

\begin{figure}[ht]
\centering
\includegraphics[width=1.\textwidth]{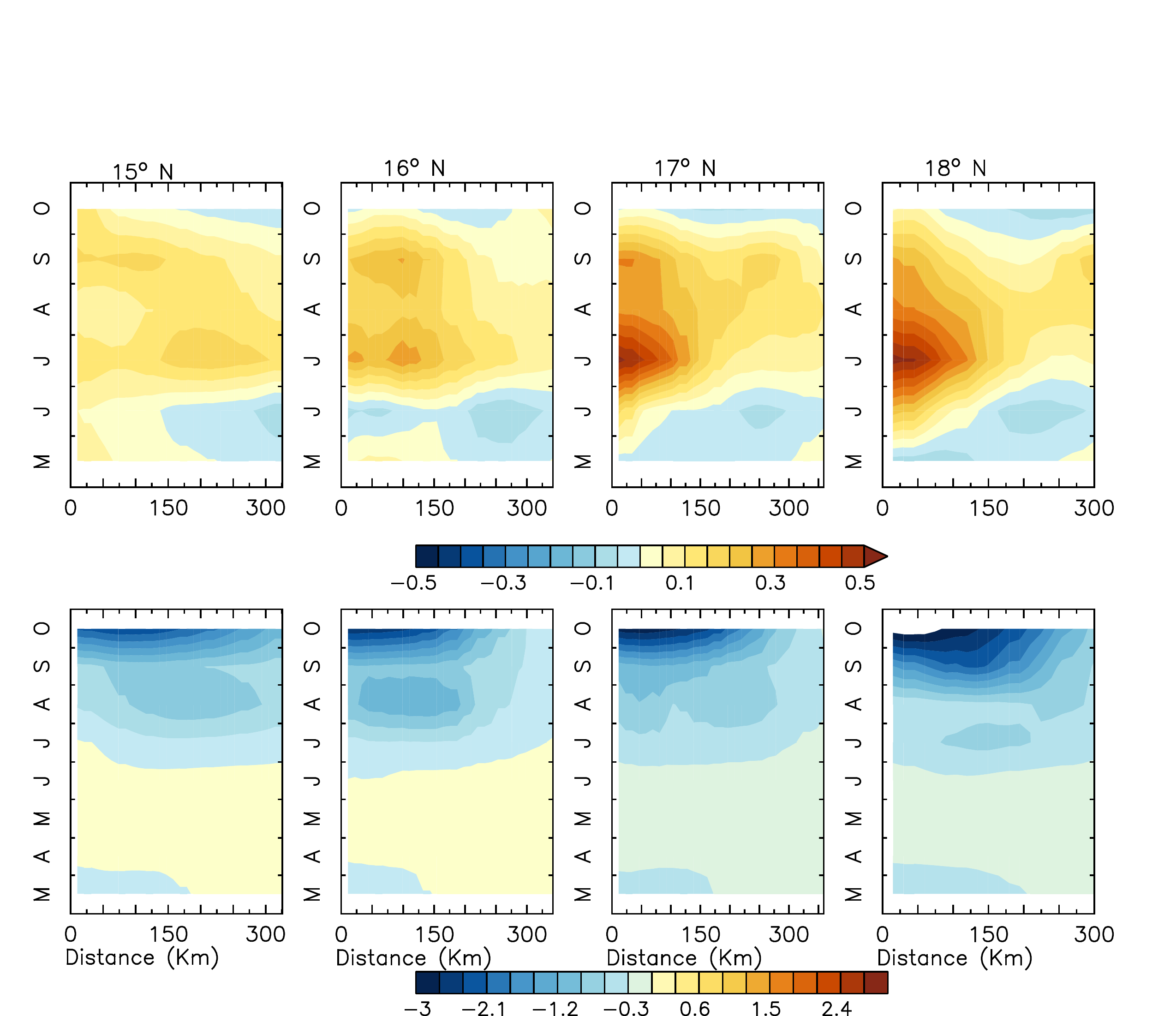} 
\caption{May to October cross-shore sections of the difference between  RIVER  and NORIVER  for SST ($^{\circ}$C, 1$^{st}$  row)  and Salinity  (psu, 2$^{nd}$  row) for the sections marked at  18$^{\circ}$N, 17$^{\circ}$N, 16$^{\circ}$N, 15$^{\circ}$N in the right panel of Figure 3.}
\label{18_15SST}
\end{figure}

\begin{figure}[ht]
\centering
\includegraphics[width=1.\textwidth]{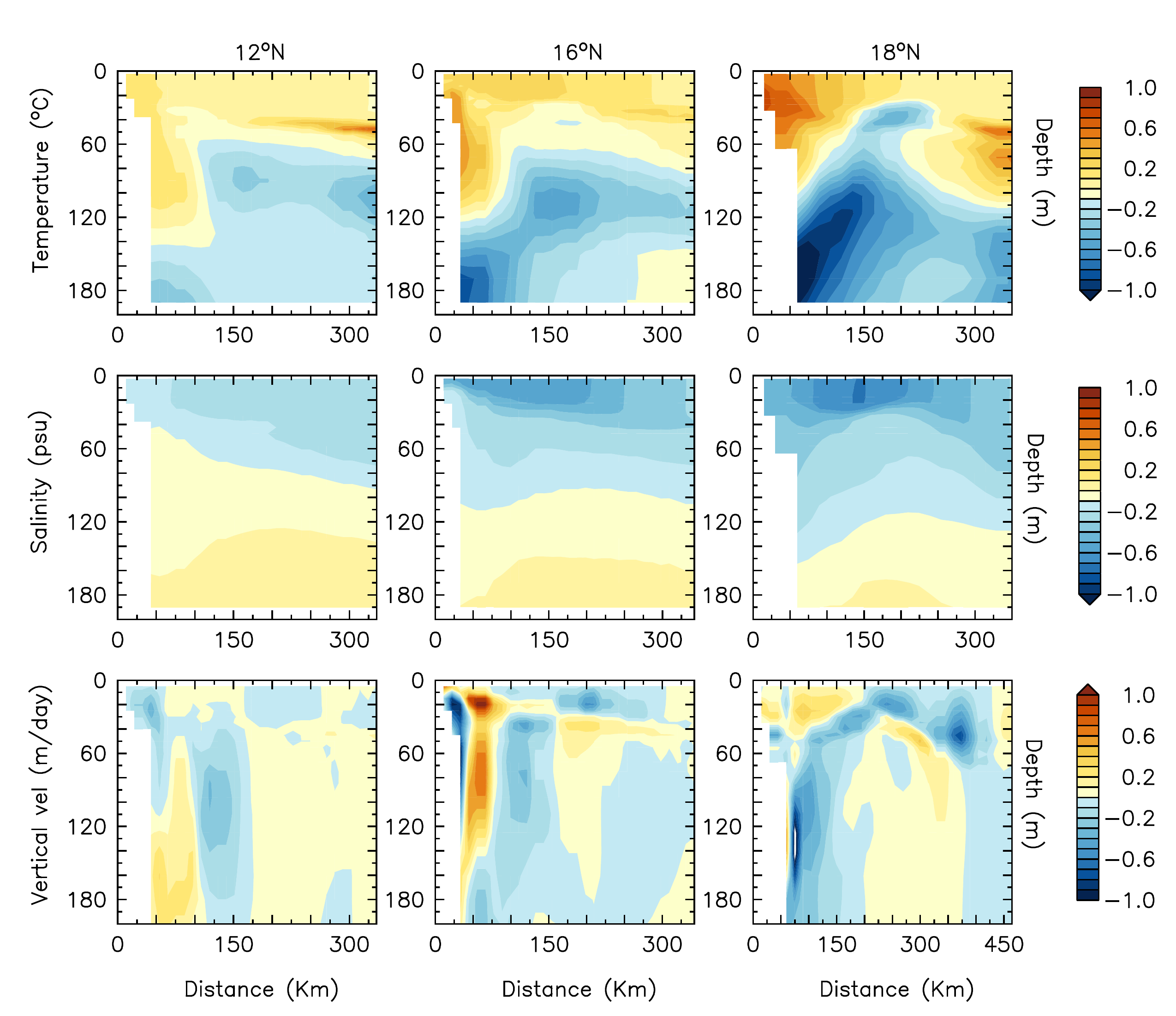} 
\caption{Vertical structure of the difference between RIVER  and NORIVER for Temperature (1$^{st}$  row), Salinity  (2$^{nd}$  row),  and Vertical velocity (3$^{rd}$ row) for the month of July for the sections marked at 12$^{\circ}$N, 16$^{\circ}$N and 18$^{\circ}$N in the right panel of Figure 3.}
\label{TSwt_diff}
\end{figure}

Low salinity water is evident in the northwestern bay which  intensifies  as the monsoon progresses.   However, towards the southwestern bay (south of 18$^{\circ}$N), the low salinity plume ($<$33 psu) becomes prominent  after August (Figure \ref{may_oct_lat_depth}).  A low salinity plume has been reported to suppress upwelling in the northwestern bay \citep{shetye1991wind, behara2016, thushara2016}. These factors prompted us to investigate  the response of upwelling to river runoff in this region.  In this section, the role of rivers in upwelling is  explained using the model experiments with the inclusion and exclusion of  river runoff into the ocean.   To identify the river impact on SST, salinity, currents, MLD, and SSH, we present the spatial maps of differences between NORIVER and RIVER simulations (Figure \ref{allvar_diff}). A lateral and vertical cross-shore section of temperature and salinity difference  is also used  to show the variations induced by rivers in the vertical and   cross-shore directions (Figure \ref{18_15SST} and Figure \ref{TSwt_diff}).



  The salinity difference between RIVER and NORIVER defines the course taken by river water in the BoB (seen as negative values in Figure \ref{allvar_diff}, 1$^{st}$ column). The river water spreads from the northern bay starting in June and has  the highest values present in the head bay during August (3 psu). The currents circulate  these waters inside the bay (vectors, 1$^{st}$ column, Figure \ref{allvar_diff}). The EICC transports it equatorward to the tip of India during the northeast monsoon (see January in Figure \ref{allvar_diff}),  while the WICC carries it northward along the west coast of India  \citep{vinayachandran2022fate}.  The EICC  strengthens equatorward (southward arrows, 1$^{st}$ column, Figure\ref{allvar_diff}) with the  inclusion of rivers  along the east coast.   A negative difference in MLD (3$^{rd}$ column,  Figure \ref{allvar_diff}) implies  shallowing  in the northwestern bay with the inclusion of rivers during August  which  spreads  southward and  shows maximum shoaling along the path of the EICC during January.   The positive SSH (4$^{th}$ column, Figure \ref{allvar_diff}) represents an increase in sea level across the entire bay and along the east coast when rivers are included.   
 
    The northwestern bay shows a positive SST difference specifying warming in the presence of rivers  ($>$0.5$^{\circ}C$),  during June which apparently shifts southward along with the transport of river water (1$^{st}$ and 2$^{nd}$ column, Figure \ref{allvar_diff}). The offshore spread of warmer SST anomalies is associated with  more robust offshore  differences in currents (vectors, 1$^{st}$ column for August in Figure \ref{allvar_diff}).  The SST and current differences  (1$^{st}$ and 2$^{nd}$ column, Figure \ref{allvar_diff}) show a clockwise recirculation of warmer water back to the eastern coast.  Thus the inclusion of rivers appears to be the reason for warming or the  suppression of upwelling-induced cooling along the east coast.  Meanwhile eastern bay cools during most of the period. In a similar experiment but by keeping the air-sea fluxes constant,  Behara and Vinayachandran (2016) noticed warming in excess of 1$^{\circ}$C in the northwestern bay and attributed it to the air-sea fluxes over a thin mixed layer owing to stratification from river and rain. The UI$_{sst}$, a measure of upwelling in the ocean   described in section 3 (dash curve, top panel, Figure \ref{UI_sst}) is higher for the NORIVER   than RIVER at all the locations.  This is an additional indication  that the weakened stratification  in the absence of rivers strengthens upwelling and  increases the cooling associated with  upwelling along the east coast of India.  Northern parts (18$^{\circ}$N) record the maximum SST difference ($>$0.5$^{\circ}$C) owing to rivers  (1$^{st}$ row, Figure \ref{18_15SST}) during June/July compared to the south. The difference in SST indicates warming  decreases offshore but  persists till October with the inclusion of rivers. The  warming caused by  rivers extends vertically (1$^{st}$ row, Figure \ref{TSwt_diff}) up to 100-120 m  near the coast. 
The negative salinity difference implies freshening in the region  which increases with the progress in monsoon and peaks during post-monsoon (in October, 2$^{nd}$ row, Figure \ref{18_15SST}) in the presence of rivers.  The highest difference occurs towards the north  ($>$ 3 psu)  gradually decreasing towards  the south.   Owing to the presence of rivers, the salinity decreases offshore ( 2$^{nd}$ row, Figure \ref{18_15SST}) over a distance of more than 150-300 km and vertically till 60 m in the south increasing to 120 m in the north (at 18$^{\circ}$N, Figure \ref{TSwt_diff}). 

 In general, warmer anomalies in the presence of  rivers are noticed over regions where salinity decreases owing to rivers.  The river discharge into the head bay during June and July is confined to the northwestern bay  (1$^{st}$ column, Figure \ref{allvar_diff}) making the region fresher and highly stratified.  The outflow of low saline water from the bay along the ECI  south of 18$^{\circ}$N starts with the southward strengthening of the  EICC (see section 3 about EICC, Figure \ref{sst_ssh_winds}) later in August/September. Fresher stratified water due to rivers  in the northern parts induces more warming and suppresses upwelling during the initial parts of the monsoon itself.  This also explains the  differential cooling during upwelling between north and south of 18$^{\circ}$ N noticed in section 3 (1$^{st}$ row and 2$^{nd}$ row, Figure \ref{sst_ssh_winds}) where the northern part warms and the southern region cool.  When rivers are switched off, the salinity increases in the northern parts of the ECI which is reflected in the temperature as it cools  more associated with upwelling (Figure \ref{18_15SST} and Figure \ref{TSwt_diff}).


 In summary, the river runoff  tends to warm the northern bay, and this warming spreads along the ECI towards the south.  Rivers cause EICC to strengthen, decrease the salinity by more than 3 psu, cause MLD to shallow along the EICC's course, and raise the sea level. In general, the southern parts of the western bay experience more upwelling-related SST cooling than the northwestern bay during the summer monsoon along the east coast (Figure \ref{may_oct_lat_depth} and Figure \ref{TSwt}). Comparatively warmer temperatures,  owing to rivers observed in the northwestern bay occurs over regions of low salinity and are observed to move southward together.  With the exclusion of rivers, the cooling from upwelling becomes stronger north of 16$^{\circ}$N where salinity is higher now (Figure \ref{18_15SST}). Without rivers, the vertical extent of upwelling increases in the western bay (100 m).  Horizontally, the impact of rivers in warming is noticed upto 300 km offshore (Figure \ref{TSwt_diff}).  
 
\section{Processes and Mechanisms}

The upper layers of the bay are highly stratified owing to low salinity caused by  the huge river discharge and enormous rainfall \citep{shetye1991wind, Vinay2002}.  
 
 In this section, we explore the processes causing the variability in upwelling in the western bay associated with the river discharge   impacted by stratification.  This is carried out by examining Brunt Vaisala frequency N$^2$ (T,S) along the ECI during the upwelling period.

 \subsection{Impact of rivers on stratification}
  The impact of salinity  on the stratification can be studied by separating the thermal (T) and haline (S) effects of N$^{2}$ \citep{maes2008ocean, maes2014seasonal}.  Computations are carried out  using the following equations. 

    \ N$^2$(T,S)   =  (g/$\rho$) d$\rho$/dz \
    
       \ N$^2$(T,S)   = N$^2$(T) + N$^2$(S) \   
       
      where \ N$^2$(T)  =  (g$\alpha$) dT/dz \

    \ N$^2$(S)         = N$^2$(T,S) - N$^2$(T) \

     where ,  $ \alpha = (-1/\rho) d\rho/dT \ $
     
\begin{figure}[ht]
\centering
\includegraphics[width=1\textwidth]{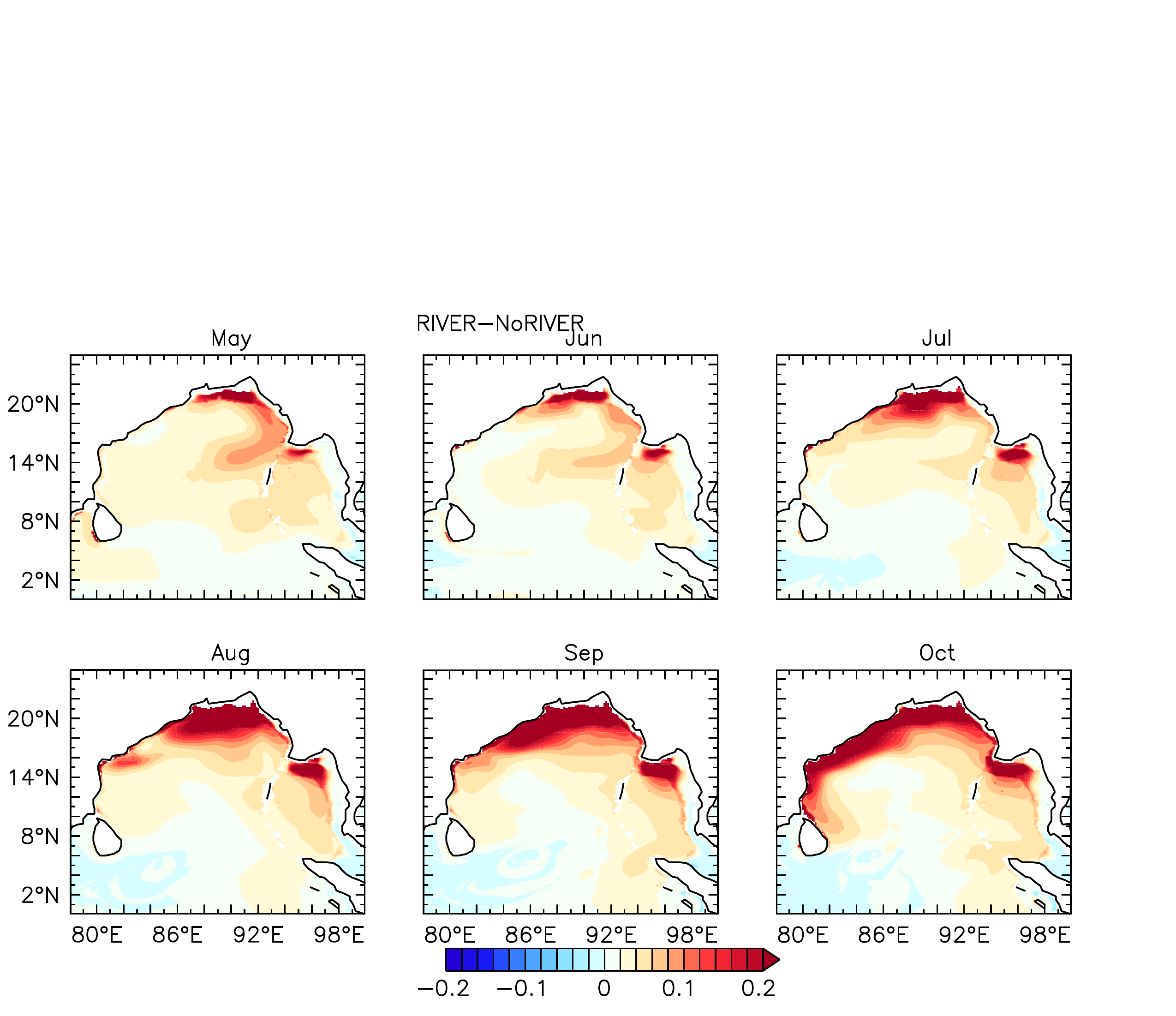} 
\caption{Difference in ocean salinity stratification (OSS, cph) between RIVER and NORIVER from May-October.}
\label{ossdiff}
\end{figure}

  T, S, and z represent temperature, salinity, and  depth respectively. $\alpha$ represents the thermal expansion coefficient, g represents gravity, and $\rho$ represents density.
  Ocean salinity stratification (OSS) is then defined as the average of  positive N$^{2}$(S) from the surface to 300 m \citep{maes2014seasonal}.   
    The difference between the OSS of RIVER and NORIVER is used from May through October to quantify the impact of rivers on salinity stratification (Figure\ref{ossdiff}). Positive difference in Figure \ref{ossdiff} indicates northern bay is stratified with respect to salinity when rivers are included.  
  Major rivers such as Ganga, Brahmaputra, Irravady, Kaveri, and Krishna discharge into the basin as the summer monsoon progresses and  OSS starts increasing  from June ($>$ 0.2 cph) spreading southwards along the ECI. By September, the highest values 
 occur till 18$^{\circ}$N, extending further south till the Sri Lankan region  in October.  Higher salinity stratification north of 18$^{\circ}$N in September corresponds to the weaker upwelling regimes in the northwestern bay.  When rivers are absent, this stratification  is weaker, which promotes the mixing of the upper oceanic layers  by wind-driven turbulence and vertical ocean processes. 
 
\begin{figure}[ht]
\centering
\includegraphics[width=1.\textwidth]{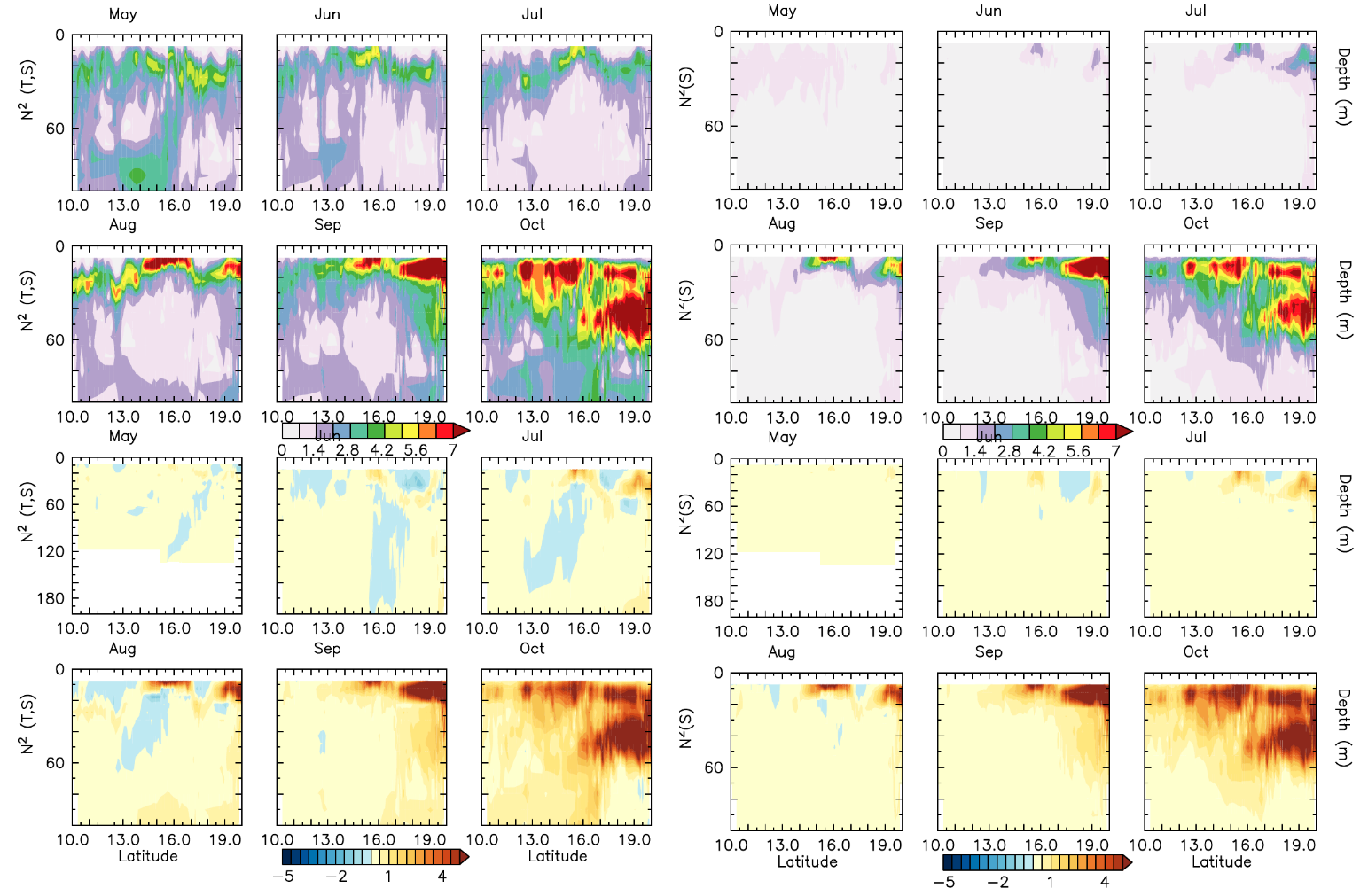} 
\caption{Left three columns. Vertical structure of N$^{2}$ (T,S) (*10$^{-4}$ /s), from the RIVER (first three columns, 1$^{st}$ and 2$^{nd}$  row) and the difference between RIVER and NORIVER (first three columns, 3$^{rd}$ and 4$^{th}$  row).  Right three columns. N$^{2}$ (S) (*10$^{-4}$ /s), from the RIVER (last three columns, 1$^{st}$ and 2$^{nd}$ row)   and  the difference between RIVER  and NORIVER  (last three columns, 3$^{rd}$ and 4$^{th}$ row). All the  plots are for data  along  the   100 m isobath  shown in the right panel of Figure 3 from May-October.}
\label{lat_depth_n_sqsalt}
\end{figure}

The variability in stratification with latitude and depth along the east coast  and the relative contribution of N$^{2}$(T,S) and N$^{2}$(S) is presented in Figure \ref{lat_depth_n_sqsalt}. As early as May and extending till July, the upper 60 m values of N$^{2}$(T,S)  are higher indicating stratification, which intensifies from August, associated  with monsoon and river discharge (1$^{st}$ three columns in 1$^{st}$ and 2$^{nd}$ row, Figure \ref{lat_depth_n_sqsalt}).  N$^{2}$(S) (Last  three columns in 1$^{st}$ and 2$^{nd}$ row, Figure \ref{lat_depth_n_sqsalt}) on the other hand starts increasing in July in the north where huge discharge from Ganga, Brahmaputra,  and major rivers occur and also between 15$^{\circ}$ and 17$^{\circ}$N owing to a river discharge from  Krishna in the model.  The highest values occur in the upper layers above 20 m  which progressively grow laterally and vertically displaying the stratification in the region induced by low salinity from  rivers.  In  October, the upper layers become highly stratified till 13$^{\circ}$N while a secondary maximum occurs north of 16$^{\circ}$N extending vertically to 60 m.  The positive values in the difference plot (last two rows, Figure \ref{lat_depth_n_sqsalt} ) show the effect of rivers on stratification.  It is very much similar for  N$^{2}$(T,S) and N$^{2}$(S), suggesting  the major contribution to stratification in this region is from rivers ((N$^{2}$(S)) where salinity plays a major role.  Higher stratification owing to rivers noticed north of 16$^{\circ}$N  slowly extends southwards  with depth after August. The major contribution for the double maximum in stratification with depth north of 16$^{\circ}$N in October is associated with river discharge.  

Therefore,  the lower salinity associated with river discharge is the reason  for the enhanced stratification during the summer monsoon and post-monsoon  in the upper layers along the ECI, particularly in the northern parts.  During the pre-monsoon to early monsoon period (1$^{st}$ three columns, Figure \ref{lat_depth_n_sqsalt}) higher stratification in the upper 60 m is contributed by temperature term as well while the salinity stratification  (last  three columns, Figure \ref{lat_depth_n_sqsalt}) increases southwards along the east coast starting from August.

     The upper layers of the western bay are generally highly stratified  depending on temperature and salinity and stratification rapidly increases from August as the salinity term starts dominating due to the increase in the 
  river discharge. The higher stratification is prominent in the upper layers till 16$^{\circ}$N in September implying that the effect of stratification is higher in the northern parts during the summer monsoon.  Stratification owing to rivers extends to the  southern parts of ECI later in October.  As observed, along the ECI,  upwelling-induced cooling increases towards the south (Figure \ref{may_oct_lat_depth}).  The local wind forcing which triggers upwelling has to work against the stratification induced by rivers in the northern parts, during the monsoon season. 
      The contribution of salinity stratification gets weaker towards the south which leads to increased cooling  to the south associated with upwelling.  This produced a differential cooling pattern between the northern and southern parts described earlier. The scenario changes when rivers are switched off, stratification weakens in the northern parts, and cooling due to upwelling becomes maximum in this region.  The cooling generated by upwelling increases by almost 0.5$^{\circ}$-1$^{\circ}$C in the northern bay, with amplitude decreasing southwards (Figure \ref{18_15SST} and Figure \ref{TSwt_diff}).  This confirms that the salinity stratification  from rivers during the summer monsoon is one reason that reduces the cooling produced by upwelling in the northwestern bay, the effect of which decreases southwards along the coast. 
The weaker stratification caused by the absence of rivers resulted in increased cooling associated with upwelling as it is easier for the subsurface water to rise to the top.  Cooler water rises from a depth of 100 m instead of 60 m (noticed in \citet{shetye1991wind}) when rivers are excluded (Figure \ref{TSwt_diff}).  The difference in stratification due to rivers being higher in the northern parts in July-September compared to southern locations suggest that rivers inhibit upwelling in the northwest bay during the summer monsoon.


\section{Summary and conclusion}
East coast of India upwelling is a seasonal phenomenon, occurring due to the seasonal monsoon winds and currents \citep{shetye1991wind,thushara2016} which push water away from the coast allowing nutrient rich and cooler water from the depth to rise to the surface layers.  BoB receives abundant rainfall and river discharges during the summer monsoon decreasing the salinity and increasing the stratification in the region. 
 Using a high-resolution OGCM configured for the global ocean, the upwelling along the ECI was studied.  The model simulated the cooling associated with upwelling along the ECI reasonable well.  The UI shows favourable conditions for upwelling beginning from February along the ECI with values ranging between 0.6 -1.2 $m^{2}s^{-1}$.  An SST based  UI index (UI$_{sst}$) and UI serve as good indicators of upwelling along the ECI. 
 The SST cooling in the western bay due to upwelling shows a differential pattern north and south of 18$^{\circ}$N with higher cooling in the south. The cooling was less in the northern parts, where salinity influence from rivers was maximum. The currents were directed poleward south of 18$^{\circ}$N and equatorward north of it in concurrence with the observed differential cooling.  
  With the progress in the monsoon, the freshwater plume, the warmer SST, and the coastal current strengthen.  The warmer waters occur over  low salinity water and  extend deeper in the northern parts of the ECI in the vertical compared to the southern parts.
   With the expansion of the freshwater plume southward and vertically, the warmer layers grew thicker equatorward.  
  In the vertical, upwelling is noticed till 60 m with downwelling occurring below it.  During upwelling, high saline water is pumped  upwards from beneath, making the upper layer more saline  before the arrival of fresher water from the northern bay. Low salinity water inhibits cooling associated with upwelling.  In order to study the impact of rivers on upwelling two experiments were carried out; one with rivers and the other without rivers. 
  
  The difference between the RIVER and NORIVER simulations  was then used to isolate the river impact on the ECI upwelling.  The salinity drops more than 3 psu in the head bay due to rivers and the river water flows along the east coast equatorward.  The presence of rivers cause the western bay to warm ( around 0.5$^{\circ}$C) and these anomalies propagate southwards along with the low saline water.  As the river water spreads southward from the northern bay, the EICC strengthens, MLD shoals and SSH rises along the east coast of India. 
  The UI$_{sst}$,  from the ocean, is higher for the NORIVER  along  the ECI implying weaker stratification in the region due to high saline waters enhances upwelling.  In general, when rivers are present upwelling related cooling increases southwards (from 27$^{\circ}$C in the north to 26$^{\circ}$C in the south). The river runoff suppresses cooling associated with upwelling, which is observed as the warming anomalies first noticed in the northwestern bay, which slowly propagates southwards along  with the low salinity water.  The exclusion of rivers causes increased cooling associated with upwelling in the northern parts while the vertical extent increases  by 40 m along the ECI close to the coast. 

 The stratification caused by salinity was examined by separating the thermal and haline effects from N$^{2}(T,S)$ and by defining the OSS as the average of N$^{2}$(S) in the upper 300 m.  The difference in OSS between the RIVER and NORIVER experiments highlights the influence of rivers on stratification. The OSS increases along the ECI with the river discharge, indicating that the region's stratification depends on river runoff.  N${^2}$(T,S) in the upper layers shows that  salinity stratification  increases from August and expands southwards  with depth.  The variability of N${^2}$(S) with depth suggests that the upper layers of the northwestern bay are highly stratified where the influence of rivers is higher.  As a result, during the summer monsoon, the rivers suppress cooling in the northwest bay than the south.   When rivers are not included,  higher salinity and weakened stratification  result in increased cooling along the path of  river flux during the summer monsoon.

         
 In summary, during the summer monsoon, rivers promote increased warming in the northwestern bay, suppressing cooling caused by upwelling through increased stratification.  Stratification by rivers emerges as the major factor suppressing ECI upwelling particularly in the northern parts. The Local winds and remote forcing from the equator dominate the sea-level cycle and associated upwelling processes in the BoB through propagating waves and eddies  \citep{SHANKAR2002}.   The equatorial wind stress forced upwelling Kelvin wave trigger upwelling in the northwest bay \citep{ray2022}.  The termination of upwelling in the northern bay after June is again associated with the arrival of downwelling coastal trapped Kelvin wave \citep{ray2022, DEY2023}.  The influence of rivers on the  coastally trapped Kelvin waves  from the head bay and along the ECI has important implications on upwelling along the ECI. Because of the fact that monthly data is not sufficient for tracking of propagating coastal waves, this part of the analysis has not been addressed here.  
 
{\bf Acknowledgements} Model simulations used in the study were carried out under the O-MASCOT project funded by INCOIS, Ministry of Earth Sciences, Govt. of India. Simulations were carried out on the MIHIR supercomputer at NCMWRF, Ministry of Earth Sciences, Govt. of India. PNV acknowledges partial financial support from J C Bose Fellowship, DST, Govt. of India, and National Supercomputing Mission, DST, Govt. of India. The model used in this study is version 5 of the modular ocean model (MOM5p1) developed at the National Oceanic and Atmospheric Geophysical Fluid Dynamics Laboratory (NOAA/GFDL).  We thank AVISO ($https://www.aviso.altimetry.fr/en/data.html$) for the SSH data and WOA18 \citep{boyer2018world} for the Salinity and Temperature data.  Ferret has been
used for data analysis and graphics. 

\bibliography{ref_upwelling.bib}
\section{Figures}

\end{document}